\documentclass[12pt]{article}
\usepackage{epsfig}
\usepackage{amssymb}
\usepackage{amsmath}
\usepackage{amsfonts}

\newcommand{\R}{\mathbb{R}}
\newcommand{\C}{\mathbb{C}}

\newcommand{\fa}{\mathfrak{a}}
\newcommand{\fb}{\mathfrak{b}}
\newcommand{\fc}{\mathfrak{c}}

\newcommand{\fn}{\mathfrak{n}}

\newcommand{\be}{\begin{equation}}
\newcommand{\ee}{\end{equation}}
\newcommand{\bea}{\begin{eqnarray}}
\newcommand{\eea}{\end{eqnarray}}
\newcommand{\nn}{\nonumber}
\newcommand{\kt}{\rangle}
\newcommand{\br}{\langle}

\newcommand{\ed}{\end{document}}

\begin{document}

\title{Explicit Realization of Pseudo-Hermitian and Quasi-Hermitian
Quantum Mechanics\\ for Two-Level Systems}
\author{\\
\underline{Ali Mostafazadeh}\thanks{Corresponding author, e-mail
address: amostafazadeh@ku.edu.tr}~ and Seher \"Oz\c{c}elik
\\
\\
Department of Mathematics, Ko\c{c} University,\\
34450 Sariyer, Istanbul, Turkey}
\date{ }
\maketitle

\begin{abstract}
We give an explicit characterization of the most general
quasi-Hermitian operator $H$, the associated metric operators
$\eta_+$, and $\eta_+$-pseudo-Hermitian operators acting in
$\C^2$. The latter represent the physical observables of a model
whose Hamiltonian and Hilbert space are respectively $H$ and
$\C^2$ endowed with the inner product defined by $\eta_+$. Our
calculations allows for a direct demonstration of the fact that
the choice of an irreducible family of observables fixes the
metric operator up to a multiplicative factor. \vspace{5mm}

\vspace{2mm}


\end{abstract}



\section{Introduction}

Recently there has been a significant interest in devising a
unitary quantum theory based on ${\cal PT}$-symmetric Hamiltonian
operators such as $H=p^2+ix^3$ that possess a real discrete
spectrum \cite{bender-1998,dorey,shin,p67}. A key ingredient that
allows for the formulation of such a quantum theory is a spectral
theorem proven in \cite{p2} (see also \cite{p3}) asserting that if
a diagonalizable operator $H$ acting in a separable Hilbert space
has a discrete spectrum, then its spectrum is real if and only if
there is a positive-definite inner product $\br\cdot,\cdot\kt_+$
on ${\cal H}$ with respect to which $H$ is Hermitian. The inner
product $\br\cdot,\cdot\kt_+$ may be conveniently expressed in
terms of a positive-definite (metric) operator $\eta_+:{\cal
H}\to{\cal H}$ according to
    \be
    \br\cdot,\cdot\kt_+=\br\cdot|\eta_+\cdot\kt,
    \label{inn}
    \ee
where $\br\cdot|\cdot\kt$ is the defining inner product of ${\cal
H}$. The Hermiticity of $H$ with respect to $\br\cdot,\cdot\kt_+$,
i.e., the condition $\br\cdot,H\cdot\kt_+=\br H\cdot,\cdot\kt_+$,
is equivalent to the $\eta_+$-pseudo-Hermiticity \cite{p1} of $H$,
namely
    \be
    H^\dagger=\eta_+ H\eta_+^{-1}.
    \label{ph}
    \ee
Another equivalent condition to the reality of the spectrum of $H$
is its quasi-Hermiticity \cite{quasi}, i.e., the existence of an
invertible operator $\rho:{\cal H}\to{\cal H}$ such that
    \be
    h:=\rho^{-1}H\rho
    \label{h}
    \ee
is Hermitian with respect to $\br\cdot|\cdot\kt$, i.e.,
$\br\cdot|h\cdot\kt=\br h\cdot|\cdot\kt$. We will call such an
operator Hermitian, i.e., use the defining inner product
$\br\cdot|\cdot\kt$ of ${\cal H}$ to determine if an operator is
Hermitian or not.

In \cite{quasi}, the authors propose a different approach to
quantum mechanics in which the inner product of the physical
Hilbert space is not a priori fixed but determined by the choice
of sufficiently many appropriate observables. In order to make
this statement more precise we first recall a few definitions.
\begin{itemize}
\item[] {\bf Definition~1}: Let ${\cal H}$ be a separable Hilbert
space. Then a set ${\cal S}=\{O_\alpha\}$ of bounded linear
operators $O_\alpha:{\cal H}\to{\cal H}$ is said to be
\emph{irreducible} if there is no proper subspace of ${\cal H}$
that is left invariant by all $O_\alpha$'s, i.e., the only
subspace ${\cal H}'$ of ${\cal H}$ satisfying the following
condition is ${\cal H}$.
    \[O_\alpha \psi'\in {\cal H}',~~~~\mbox{for all}~~~~
    \psi'\in{\cal H}'~~~~{\rm and~all}~~O_\alpha\in
    {\cal S}.\]

\item[] {\bf Definition~2}: Let ${\cal H}$ be a separable Hilbert
space. Then a set ${\cal S}=\{O_\alpha\}$ of quasi-Hermitian
linear operators $O_\alpha:{\cal H}\to{\cal H}$ is said to be
\emph{compatible} if there is an invertible bounded operator
$\rho:{\cal H}\to{\cal H}$ such that $\rho^{-1}O_\alpha\rho$ is
Hermitian for all $O_\alpha\in{\cal S}$.
\end{itemize}
The condition that $\rho^{-1}O_\alpha\rho$ is Hermitian is
equivalent to the existence of a positive-definite (metric)
operator $\eta_+$ such that $O_\alpha$ is
$\eta_+$-pseudo-Hermitian \cite{p2}. Thus, ${\cal S}=\{O_\alpha\}$
is a compatible set if and only if all $O_\alpha$ are
$\eta_+$-pseudo-Hermitian for some metric operator $\eta_+$.

We can express the main result of \cite{quasi} as:
\begin{itemize}
\item[] {\bf Theorem}: Up to a multiplicative factor there is a
unique positive-definite (metric) operator $\eta_+$ such that all
the elements of a compatible irreducible set ${\cal
S}=\{O_\alpha\}$ of operators is $\eta_+$-pseudo-Hermitian.
Equivalently, there is a (positive-definite) inner product
$\br\cdot,\cdot\kt_+$ on ${\cal H}$ such that all $O_\alpha$ are
Hermitian with respect to $\br\cdot,\cdot\kt_+$, and this inner
product is unique up to a trivial multiplicative factor.
\end{itemize}

The purpose of this article is to conduct a thorough investigation
of the implementation of the above-mentioned developments to the
simplest nontrivial class of quantum systems, namely the two-level
systems for which ${\cal H}$ is $\C^2$ endowed with the Euclidean
inner product. In particular, we
\begin{itemize}
\item compute the most general form of quasi-Hermitian operators,
\item find the explicit form of the most general metric operator
$\eta_+$ that renders a given quasi-Hermitian operator
$\eta_+$-pseudo-Hermitian, \item determine the class of all
quasi-Hermitian operators $O$ that together with the given one
($H$) form a compatible irreducible set, and \item show by an
explicit calculation how the choice of such an operator fixes
$\eta_+$ up to a scale factor. \end{itemize}

In the remainder of this paper ${\cal H}$ will denote $\C^2$
endowed with the Euclidean inner product, the elements of ${\cal
H}$ will be represented by column vectors, and the linear
operators $L:{\cal H}\to{\cal H}$ will be identified with their
matrix representations in the standard basis of $\C^2$. In
particular, both the identity operator acting in $\C^2$ and the
$2\times 2$ identity matrix will be denoted by $I$.

\section{Quasi-Hermitian Operators and the Associated\\ Metric
Operators}

It is clear that every quasi-Hermitian operator $H$ is necessarily
diagonalizable and has a real spectrum. These conditions hold if
and only if $H$ is $\eta_+$-pseudo-Hermitian for a
positive-definite (metric) operator $\eta_+$, \cite{p2,p3}.
Furthermore, $H$ is $\eta_+$-pseudo-Hermitian if and only if its
traceless part $H_0:=H-\frac{1}{2}{\rm tr}(H)I$ is
$\eta_+$-pseudo-Hermitian.\footnote{This is because the
eigenvalues of $H$ and hence its trace are real.} As a result we
can confine our attention to traceless operators:
    \be
    H_0=\left(\begin{array}{cc}
    \fa&\fb\\
    \fc&-\fa\end{array}\right)~~~~{\rm with}~~~~\fa,\fb,\fc\in\C.
    \label{H}
    \ee
Next, we use the conditions that $H_0$ is diagonalizable and its
eigenvalues are real and have opposite sign to infer that $\det
H_0=-(\fa^2+\fb\fc)$ is real and non-positive. This in turn
implies that the eigenvalues of $H_0$ are given by $\pm E$ where
$E:=\sqrt{\fa^2+\fb\fc}$. The case $E=0$ corresponds to the
trivial degenerate case where $H_0$ is the zero operator.

The converse of the above argument also holds, i.e., the condition
$\fa^2+\fb\fc\in\R^+$ ensures that $H$ is a nonzero
quasi-Hermitian operator. Relaxing the tracelessness condition, we
have the following form for the most general quasi-Hermitian
operator acting in ${\cal H}$.
    \be
    H=H_0+qI=\left(\begin{array}{cc}
    q+\fa&\fb\\
    \fc&q-\fa\end{array}\right),
    \label{Q}
    \ee
where $q\in\R$, $\fa,\fb,\fc\in\C$, and
$\fa^2+\fb\fc\in[0,\infty)$.

An alternative parametrization of $H_0$ that simplifies its
diagonalization is \cite{angle}
    \be
    H_0=E\left(\begin{array}{cc}
    \cos\theta&e^{-i\varphi}\sin\theta\\
    e^{i\varphi}\sin\theta&-\cos\theta\end{array}\right),
    \label{H=}
    \ee
where $E\in[0,\infty)$, $\theta,\varphi\in\C$,
$\Re(\theta)\in[0,\pi)$ and $\Re(\varphi)\in[0,2\pi)$.\footnote{
$\Re$ and $\Im$ stand for the real and imaginary parts of their
argument.} The adjoint $H_0^\dagger$ of $H_0$ and any linearly
independent pair of eigenvectors $|\phi_n\kt$ of $H_0^\dagger$ are
given by
    \bea
    &&H_0^\dagger=E\left(\begin{array}{cc}
    \cos\theta^*&e^{-i\varphi^*}\sin\theta^*\\
    e^{i\varphi^*}\sin\theta^*&-\cos\theta^*\end{array}\right),
    \label{H-dag}\\
    &&|\phi_1\kt={\fn}_1\left(\begin{array}{c}
    \cos\mbox{$\frac{\theta^*}{2}$}\\
    e^{i\varphi^*}\sin\mbox{$\frac{\theta^*}{2}$}\end{array}\right),
    ~~~~~~|\phi_2\kt={\fn}_2\left(\begin{array}{c}
    \sin\mbox{$\frac{\theta^*}{2}$}\\
    -e^{i\varphi^*}\cos\mbox{$\frac{\theta^*}{2}$}\end{array}\right),
    \eea
where $\fn_1,\fn_2\in\C-\{0\}$ are arbitrary.

The most general positive-definite metric operator $\eta_+$ that
renders $H_0$ (and hence $H$) $\eta_+$-pseudo-Hermitian is given
by \cite{p4,jmp-2003}
    \be
    \eta_+=\sum_{n=1}^2|\phi_n\kt\br\phi_n|=k
    \left(\begin{array}{cc}
    au+b & e^{-i\varphi}(u \zeta-\zeta^*)\\
    e^{i\varphi^*}(u\zeta^*-\zeta) & e^{i(\varphi^*-\varphi)}
    (a+b u)\end{array}\right),
    \label{eta}
    \ee
where
    \bea
    &&a:=|\cos\mbox{$\frac{\theta}{2}$}|^2,~~~
    b:=|\sin\mbox{$\frac{\theta}{2}$}|^2,~~~
    \zeta:=\sin\mbox{$\frac{\theta}{2}$}
    \cos\mbox{$\frac{\theta^*}{2}$}~,
    \label{param1}\\
    &&k:=|\fn_2|^2,~~~~~~u:=|\fn_1/\fn_2|^2.\nn
    \eea
Note that the parameters $a,b,\zeta$ are determined by $\theta$
and $\varphi$ that fix $H_0$ while $k$ and $u$ characterize the
freedom of the choice of $\eta_+$. In particular, changing $k$
corresponds to a trivial scaling of $\eta_+$ and the associated
inner product (\ref{inn}).

For later use we introduce the set ${\cal U}_{_H}$ consisting of
all the metric operators (\ref{eta}) such that $H$ is
$\eta_+$-pseudo-Hermitian.

\section{Irreducible Sets of Compatible Quasi-Hermitian\\ Operators}

Given a linear operator acting in $\C^2$, the eigenspaces
associated with each eigenvalue is an invariant subspace of the
operator. This implies that a pair of (diagonalizable
quasi-Hermitian) operators have a common proper invariant
subspace, if they share an eigenvector. In other words, they form
an irreducible set, if they have no common eigenvectors. In light
of this observation, to construct an irreducible set of compatible
quasi-Hermitian operators one needs to supplement a given
quasi-Hermitian operator $H$ with a linear operator $H'$ that is
$\eta_+$-pseudo-Hermitian for some $\eta_+\in {\cal U}_{_H}$ and
do not share any eigenvector with $H$.

We shall first consider the problem of the characterization of a
general quasi-Hermitian operator that is compatible with $H$. In
other words, we shall construct the most general quasi-Hermitian
operator, $H'=H_0'+q'I$ with
    \be
    H_0'=\left(\begin{array}{cc}
    \fa'&\fb'\\
    \fc'&-\fa'\end{array}\right),
    \label{H-prime}
    \ee
$\fa',\fb',\fc'\in\C$, ${\fa'}^2+\fb'\fc'\in[0,\infty)$ and
$q'\in\R$, such that ${H'}^\dagger=\eta_+H'\eta_+^{-1}$ for a
metric operator $\eta_+$ of the form (\ref{eta}). This in turn
implies ${H_0'}^\dagger=\eta_+H_0'\eta_+^{-1}$ or alternatively
    \be
    {H'_0}^\dagger\eta_+=\eta_+ H'_0.
    \label{ph-prime}
    \ee
This relation leads to four complex equations for the three
unknowns $\fa',\fb'$ and $\fc'$. These equations that involve the
fixed complex constants $\theta,\varphi$ (alternatively
$a,b,\zeta$) and the free real positive parameter $u$ are not
independent. They can be reduced to the following three simpler
equations.
    \bea
    \Im(\lambda\fb')&=& r\Im(\fa'),
    \label{eq1}\\
    \Im(\lambda{\fc'}^*)&=& s\Im(\fa'),
    \label{eq2}\\
    s\fb'-r{\fc'}^*&=&2\lambda^*\Re(\fa'),
    \label{eq3}
    \eea
where
    \be
    \lambda:=e^{i\varphi^*}(u\zeta^*-\zeta),~~~~
    r:=e^{2\Im(\varphi)}(a+bu),~~~~
    s:=au+b.
    \label{param2}
    \ee
The parameters $\fa',\fb'$, and $\fc'$ entering the expression for
$H_0'$ are furthermore subject to the constraint
${\fa'}^2+\fb'\fc'\in[0,\infty)$, i.e.,
    \bea
    \Re({\fa'}^2+\fb'\fc')&\geq& 0,
    \label{condi-1}\\
    \Im({\fa'}^2+\fb'\fc')&=& 0.
    \label{condi-2}
    \eea
Note, however, that in view of the $\eta_+$-pseudo-Hermiticity of
$H_0'$ and the positive-definiteness of $\eta_+$, $H_0'$ is
Hermitian with respect to the positive-definite inner
product~(\ref{inn}). This in turn implies that it is
diagonalizable and has a real spectrum. As a result, we expect
conditions (\ref{condi-1}) and (\ref{condi-2}) not to lead to any
further restrictions on the possible values of $\fa',\fb'$, and
$\fc'$. In the following we shall first solve
(\ref{eq1})--(\ref{eq3}) and check by explicit calculation that
indeed (\ref{condi-1}) and (\ref{condi-2}) are automatically
satisfied.

Before exploring (\ref{eq1})--(\ref{eq3}), we note the following
useful identity:
    \be
    \zeta=
    \frac{1}{2}\left(\sin[\Re(\theta)]+i\sinh[\Im(\theta)]\right).
    \label{zeta=}
    \ee
which follows from (\ref{param1}) upon the application of Euler's
formula. We can use (\ref{zeta=}) to infer the equivalence of the
conditions: $\theta=0$ and $\zeta=0$.\footnote{Recall that
$\Re(\theta)\in[0,\pi)$.} Another consequence of (\ref{zeta=}) is
that for all nonzero values of $\zeta$, $\lambda=0$ if and only if
$\zeta^*/\zeta$ is real and equal to $u$. In this case, $\zeta$
and hence $\theta$ must be real and $u=1$.

Now, we are in a position to analyze (\ref{eq1})--(\ref{eq3}),
(\ref{condi-1}) and (\ref{condi-2}). We do this by considering the
following two cases separately.
\begin{itemize}
\item[]\textbf{Case 1)} $\lambda=0$: This is equivalent to setting
(1) $\zeta=\theta=0$, or (2) $\zeta,\theta\in\R$ and $u=1$. In
this case (\ref{eq1})--(\ref{eq3}) yield
    \be
    \Im(\fa')=0,~~~~~~\fc'=r^{-1}s\,{\fb'}^*=
    \left(\frac{au+b}{a+bu}\right)e^{-2\Im(\varphi)}{\fb'}^*,
    \label{sol-1}
    \ee
where $\Re(\fa')=\fa'$ and $\fb'$ are respectively real and
complex free parameters. In view of (\ref{sol-1}), it is not
difficult to check that indeed (\ref{condi-1}) and (\ref{condi-2})
are automatically satisfied.

\item[]\textbf{Case 2)} $\lambda\neq0$ where either (3) $u\neq 1$
and $\zeta,\theta\neq 0$, or (4) $\zeta$ and $\theta$ are not
real. In this case we can simplify (\ref{eq3}) by multiplying its
both sides by $\lambda$. Then the imaginary part of the resulting
equation is automatically satisfied by virtue of (\ref{eq1}) and
(\ref{eq2}), and its real part yields
    \be
    s\Re(\lambda\fb')-r\Re(\lambda{\fc'}^*)=2|\lambda|^2\Re(\fa').
    \label{eq4}
    \ee
We can solve (\ref{eq1}), (\ref{eq2}), and (\ref{eq4}) to obtain:
    \bea
    \fb'&=&\lambda^{-1}[w+i r\Im(\fa')]=\frac{
    e^{-i\varphi^*}w+ie^{-i\varphi}(a+bu)\Im(\fa')}{u\zeta^*-\zeta},
    \label{sol-21}\\
    \fc'&=&(r\lambda^*)^{-1}[s w-2|\lambda|^2\Re(\fa')-irs\Im(\fa')]
    \nn\\&=&
    \frac{e^{i\varphi}[(au+b)e^{-2\Im(\varphi)}w-2|u\zeta-\zeta^*|^2
    \Re(\fa')-i(a+bu)(au+b)\Im(\fa')]}{(u\zeta-\zeta^*)(a+bu)},
    \label{sol-22}
    \eea
where $w\in\R$ and $\fa'\in\C$ are arbitrary. Next, we impose the
constraints (\ref{condi-1}) and (\ref{condi-2}). In view of the
fact that $\lambda\neq 0$, (\ref{condi-2}) is equivalent to
    \be
    \Im[|\lambda|^2{\fa'}^2+(\lambda\fb')(\lambda{\fc'}^*)^*]=0.
    \label{condi-2n}
    \ee
Expressing the left-hand side of this relation in terms of the
real and imaginary parts of $\lambda\fb'$ and $\lambda{\fc'}^*$
and making use of (\ref{eq1}), (\ref{eq2}) and (\ref{eq4}), we
have checked that (\ref{condi-2n}) is automatically satisfied.
Finally imposing (\ref{condi-1}) yields
    \be
    |\lambda|^2\left[\Re(\fa')^2-\Im(\fa')^2-2r^{-1}w\Re(\fa')\right]
    +rs\Im(\fa')^2+r^{-1}sw^2\geq 0.
    \label{cons-1}
    \ee
We shall next prove that this inequality is also satisfied for all
$w\in\R$ and $\fa'\in\C$ irrespective of values of
$\theta,\varphi$ and $u$. To do this we first view (\ref{cons-1})
as a quadratic polynomial in $\Im(\fa')$. This polynomial will be
non-negative if we can prove that it does not have two distinct
real roots. This is equivalent to the condition:
    \be
    D\geq 0,
    \label{D}
    \ee
where
    \bea
    D&:=&\frac{|\lambda|^2[\Re(\fa')^2-2r^{-1}w\Re(\fa')]+r^{-1}sw^2}{
    rs-|\lambda|^2}\nn\\
    &=&\frac{|\lambda|^{2} \left\{ [\Re(\fa')-r^{-1}w]^2+
    r^{-2}w^2(rs-|\lambda|^{2})\right\}}{
    rs-|\lambda|^{2}}.
    \label{D=}
    \eea
As seen from this relation, in order to show that $D\geq 0$ we
just need to check that $rs-|\lambda|^{2}$ is non-negative.
Employing (\ref{param1}), (\ref{param2}), and the elementary
trigonometric identities
    \[\cos^2\mbox{$\frac{\theta}{2}$}=\frac{1+\cos\theta}{2},~~~~
    \sin^2\mbox{$\frac{\theta}{2}$}=\frac{1-\cos\theta}{2},\]
we have
    \bea
    rs-|\lambda|^{2}&=&e^{2\Im(\varphi)}
    [(a+bu)(au+b)-|u\zeta-\zeta^*|^2]\nn\\
    &=&e^{2\Im(\varphi)} u \left(
    |\cos\mbox{$\frac{\theta}{2}$}|^2+
    |\sin\mbox{$\frac{\theta}{2}$}|^2+2\Re[
    \cos^2\mbox{$\frac{\theta}{2}$}
    \sin^2\mbox{$\frac{\theta}{2}$}^*]\right)\nn\\
    &=&e^{2\Im(\varphi)} u.
    \label{condi-5}
    \eea
Hence $rs-|\lambda|^{2}>0$, and (\ref{D}) and (\ref{cons-1}) are
always satisfied.
\end{itemize}

In summary, the traceless quasi-Hermitian operators $H_0'$ that
together with $H_0$ (respectively $H$) form a compatible set
involve, in addition to the complex parameters $\theta,\varphi$ of
$H_0$ and the real positive parameter $u$ of $\eta_+$, three free
real parameters. For Case~1, these can be taken as
$\Re(\fa'),\Re(\fb')$ and $\Im(\fa')$. For Case~2, one can choose
$\Re(\fa'),\Im(\fa')$ and $w$ (or alternatively $\Re(\fb')$).

Next, we return to the discussion of the irreducibility of a set
of compatible quasi-Hermitian operators. As we mentioned above two
operators $H$ and $H'$ form an irreducible set, if they lack a
common eigenvector. This is equivalent to the condition
    \be
    \det([H,H'])\neq 0,
    \label{det}
    \ee
where $[\cdot,\cdot]$ denotes the commutator. It is easy to see
that (\ref{det}) is equivalent to
    \be
    \det([H_0,H'_0])\neq 0,
    \label{det-H}
    \ee
where $H_0$ and $H_0'$ are the traceless parts of $H$ and $H'$,
respectively. Substituting (\ref{H}) and (\ref{H-prime}) in
(\ref{det-H}), we find
    \be
    (\fb\fc'-\fc\fb')^2-4(\fa\fb'-\fb\fa')(\fa\fc'-\fc\fa')\neq 0.
    \label{irr}
    \ee

Let ${\cal C}_{_H}$ denote the moduli space of the quasi-Hermitian
operator $H'$ that are compatible with $H$. Every point in ${\cal
C}_{_H}$ is parameterized by the free parameters $q'\in\R$ that
equals ${\rm tr}(H')/2$, $u\in\R^+$ that enters in the expression
for the allowed metric operators $\eta_+$, and the three free real
variables: $\Re(\fa'),\Re(\fb')$ and $\Im(\fa')$ for Case~1 and
$\Re(\fa'),\Im(\fa')$ and $w$ (or alternatively $\Re(\fb')$) for
Case~2. The moduli space ${\cal M}_{_H}$ of the quasi-Hermitian
operators $H'$ that are compatible with $H$ and together with $H$
constitute an irreducible set is the subset of ${\cal C}_{_H}$
that exclude the values of the latter three variables for which
the left-hand side of (\ref{irr}) vanishes. ${\cal C}_{_H}-{\cal
M}_{_H}$ is a three dimensional subspace of ${\cal C}_{_H}$. Hence
a generic choice for $H'\in{\cal C}_{_H}$ will belong to ${\cal
M}_{_H}$; $H$ and $H'$ will form an irreducible set.

Finally, we consider fixing an element $H'$ of ${\cal
M}_{_H}-\{H\}$, i.e., selecting particular values for $q'$,
$\fa'$, $\fb'$, $\fc'$ that fulfil (\ref{eq1})--(\ref{eq3}) and
(\ref{irr}). We wish to explain why in this case the value of $u$
is uniquely determined and hence $\eta_+$ is fixed up to the
scaling factor $k$. Consider Case~1 above. If $\theta\neq 0$, then
$u=1$ necessarily. If $\theta=0$, then we can solve for $u$ in the
second equation in (\ref{sol-1}). The latter is linear in $u$,
hence its solution is unique. Next, consider Case~2. Then we can
similarly solve (\ref{sol-21}) for $u$. Again this equation is
linear in $u$ and its solution is unique.

\section{Concluding Remarks}

In this paper we have provided an explicit characterization of the
most general quasi-Hermitian operator $H$, the associated metric
operators $\eta_+$, and $\eta_+$-pseudo-Hermitian operators acting
in $\C^2$. We have derived a quantitative condition that singles
out the operators $H'$ that together with $H$ form an irreducible
set of compatible quasi-Hermitian operators, and demonstrated how
a choice of $H$ and $H'$ fixes the metric operator up to a scale
factor.

In pseudo-Hermitian quantum mechanics \cite{jpa-2004b}, a quantum
system is constructed by choosing a Hamiltonian operator $H$ (from
among the set of quasi-Hermitian operators) and a metric operator
$\eta_+$ that renders the Hamiltonian $\eta_+$-pseudo-Hermitian.
The latter specifies the physical Hilbert space of the system. The
physical observables are represented by $\eta_+$-pseudo-Hermitian
operators $H'$. In quasi-Hermitian quantum mechanics \cite{quasi},
instead of choosing $H$ and $\eta_+$ one chooses $H$ and another
element of the space ${\cal M}_{_H}$ of all quasi-Hermitian
operators that together with $H$ form an irreducible set of
compatible operators.

As we demonstrated, by explicit calculations for a general
two-level quantum system, employing quasi-Hermitian quantum
mechanics requires carrying out all the constructions of the
pseudo-Hermitian quantum mechanics. In this sense, the former is
not more practical than the latter.

It is occasionally argued that quasi-Hermitian quantum mechanics
is more physically relevant because it involves fixing the Hilbert
space after choosing a set of physical observables. We wish to
emphasize that the physical interpretation of an operator cannot
be achieved before fixing the structure of the Hilbert space it
acts in. Therefore, as far as the physical aspects of both pseudo-
and quasi-Hermitian formulations of quantum mechanics are
concerned, they are on equal grounds.

\ed